\documentstyle[sprocl]{article}

\def\bea{\begin{eqnarray}}
\def\eea{\end{eqnarray}}
\newcommand{\be}{\begin{eqnarray}}
\newcommand{\ee}{\end{eqnarray}}
\newcommand{\ben}{\begin{eqnarray*}}
\newcommand{\een}{\end{eqnarray*}}
\newcommand{\la}{\langle}
\newcommand{\ra}{\rangle}
\newcommand{\lla}{\left\langle}
\newcommand{\rra}{\right\rangle}

\def\diffn#1      {\Delta^{-}_{#1}}
\def\mb#1       {\mbox{\boldmath $#1$}}

\def\preprints{
\vspace{-14ex}
{\small
\begin{tabbing}
\` {\sl hep-lat/9808053} \\
    \\
\` LSUHE No. 272--1998 \\
\` August 1998 \\
\end{tabbing} 
}
\vspace*{0.1in}
}

\def\hayfoot
{\footnote{presented by R.H. at 
the Fourth workshop on Quantum Chromodynamics 1-6 June 1998,
The American University of Paris}
}

\begin{document}

\title{
\preprints
Electric and magnetic U(1) currents in lattice confinement studies}

\author{Alistair Hart, Richard W. Haymaker%
\hayfoot%
~and 
Yuji Sasai\footnote{permanent address: Oshima National College of Maritime 
Technology, 1091-1, Oshima -cho, Oshima-gun, Yamaguchi 743-21, Japan}
}

\address{Department of Physics and Astronomy, 
Louisiana State University,\\
Baton Rouge, Louisiana 70803,  USA}


\maketitle

\abstracts{Making use of an Ehrenfest-Maxwell relation we
show that in Abelian projected SU(2), in the maximal 
Abelian gauge, the dynamical 
electric charge density generated by the coset fields, gauge
fixing and ghosts  shows antiscreening as in the case of 
the non-Abelian charge.}   

\section{Introduction}

Lattice studies based on Abelian projection have had 
considerable success identifying
the dynamical variables relevant to the physics of
quark confinement. There is no definitive way as yet
of choosing the optimum variables, but in the 
maximal Abelian gauge 
\cite{thooft81,kronfeld87a}
the U(1) fields remaining after Abelian projection
produce a heavy quark potential that continues to rise
linearly 
\cite{suzuki90}.
Further the string tension is almost, but not exactly,
equal to the full SU(2) quantity; 92\% in a recent study
at $\beta = 2.5115$
\cite{bali96}.

This suggests that we may be close to identifying an
underlying principle governing confinement.
All elements of a dual superconducting vacuum appear to
be present 
\cite{mandelstam76,thooft81};
in the maximal Abelian gauge  
magnetic monopoles reproduce nearly all of the U(1) string
tension
\cite{stack94,bali96}.
The spontaneous breaking to the U(1) gauge symmetry 
is signalled by the
non-zero vacuum expectation value of monopole operator
\cite{chernodub96,digiacomo98}.
The profile of the electric field and the 
persistent magnetic monopole currents 
in the vortex between quark and antiquark  
are well described by an
effective theory, the Ginzburg--Landau, or equivalently a Higgs
theory giving a London penetration depth and Ginzburg--Landau 
coherence length
\cite{singh93,bali98}.

Central to finding the effective theory is the definition of
the field strength operator in the Abelian projected theory,
entering not only in the vortex profiles but also in the
formula for the monopole operator. All definitions should be
equivalent in the continuum limit, 
but use of the appropriate lattice
expression should lead to a minimization of discretization
errors.

In Ref.
\cite{dhh98}
we exploit lattice symmetries to derive such an
operator that satisfies Ehrenfest relations; Maxwell's 
equations for ensemble averages irrespective of
lattice artifacts. 

The charged coset fields are normally discarded in Abelian
projection, as are the ghost fields arising from the 
gauge fixing procedure. Since the remainder of the SU(2)
infrared physics must arise from these, an understanding of
their r{\^o}le is central to completing the picture of full
SU(2) confinement.  In the maximal Abelian gauge a 
localised cloud of like polarity charge is
induced in the vacuum in the vicinity of a source, producing
an effect reminiscent of the antiscreening of charge in 
{\sc QCD}. In other gauges studied, the analogous current
is weaker, and acts to {\em screen} the source
\cite{bdh}. 
(This a tentative result, however, without the benefit of 
the refined definition of flux.)

\section{Maxwell Ehrenfest relation}

We first introduce and review the method due to Zach et al.
\cite{zach95}
in pure U(1) theories.  Consider a shift in a U(1) 
link angle, 
$\theta_\nu(x_0) \to \theta_\nu(x_0) + \theta^s(x_0)$,
in the partition function 
containing a Wilson loop source term
\ben
Z_W(\{\theta^s\}) = \int [d (\theta_{\nu} + \theta^s) ] \;\;
\sin \theta_W  \;\;e^{\beta \sum \cos \theta_{\mu \nu}}.
\een
Since the Haar measure is invariant under this shift, $Z_W$ is constant
in these variables.  Absorbing the shift into the integration variable
and taking the derivative gives
\ben
\frac{\partial}{\partial \theta^s(x_0)} Z_W = 
 \int [d \theta] \;
(\cos \theta_W - \sin \theta_W \;\beta \Delta_{\mu} \sin \theta_{\mu \nu})
\;e^{\beta \sum \cos \theta_{\mu \nu}} = 0.
\een
This can be cast into the form
\ben
\la \Delta_{\mu} F_{\mu \nu} \ra_W  \equiv 
\frac{\la \sin \theta_W \;\;
\Delta_{\mu}\frac{1}{e} \sin \theta_{\mu \nu} \ra }
{\la \cos \theta_W \ra} =  e \delta_{x, x_W} = J_{\nu}.
\een
We use the term Ehrenfest-Maxwell relation because 
it is the expectation value of what is normally a classical extremum
of the path integral --- an Euler--Lagrange equation.  If we define 
flux using  $\sin \theta_{\mu \nu}$ instead of
$ \theta_{\mu \nu}$ for example, then we get a precise 
lattice definition of current.

Before addressing the full problem we first generalize from
U(1) to SU(2) without the complication of gauge fixing,
with a shift
$U_\mu(x_0) \to U_\mu(x_0) U^s(x_0)$
\ben
Z_W(\{U^s\}) &=& \int [d (UU^s)]\;\; W_3(U)\;\; e^{\beta S(U)}
; \;\;\;\;
W_3 \equiv \frac{1}{2}Tr (U^{\dagger}U^{\dagger} U U i \sigma_3).
\een
The size of the source is irrelevant so we choose 
it to be the simplest case, i.e. a plaquette.
We choose the shift to be in the $3$ direction
\ben
\frac{d}{d \epsilon_{\mu}(x_0)}Z_W = 0; \;\;\;\;
U^s(x_0) = \left(1 - \frac{i}{2}\epsilon_{3}(x_0) \sigma_3 \right),
\een
giving the Ehrenfest relation
\ben
\beta \frac{\la W_3 S_{\mu}  \ra}{\la W  \ra} =
 \delta_{x, x_W} ; \;\;\;\; 
 W \equiv \frac{1}{2}Tr (U^{\dagger}U^{\dagger} U U ).
\een
The notation $(S)_{\mu}$ denotes an $\epsilon$ derivative
\cite{dhh98}.

For $\beta=2.5$,   $\beta \la W_3 (S)_{\mu} \ra  = 0.0815(2)$,   and  
$\la W  \ra  = 0.0818(1)$, and the difference $= 0.0003(3)$;
i.e. zero within statistical errors.  

To cast this into the form of Maxwell's equation we
decompose the link into diagonal $D_{\mu}$ and off-diagonal parts
$O_{\mu}$
\ben
U_{\mu}(x) = D_{\mu}(x) + O_{\mu}(x).
\een
We then group terms involving the diagonal part 
into $div E$    and 
group all terms having at least one factor of the off-diagonal
link into the current.  
\ben
\left[\beta \la (S)_{\mu}  \ra_W \right]_{U = D} =  
\frac{1}{e} \la  \Delta_{\mu} F_{\mu \nu} \ra_W  
; \;\;\;\;
\la  \cdots  \ra_W \equiv \frac{\la W_3 \cdots\ra }{\la W \ra},
\een
giving the final form of the Ehrenfest relation
\ben
\la \Delta_{\mu} F_{\mu \nu} \ra_W 
= \la J_{\nu}^{dyn.} \ra_W  + J_{\nu}^{static}; \;\;\;\;
\delta_{x, x_W} = \frac{1}{e} J_{\nu}^{static}. 
\een
This then tells us how to choose a lattice definition of field 
strength that satisfies an Ehrenfest relation:
\ben
F_{\mu \nu} = 
\frac{1}{e} \frac{1}{2}
Tr(D^{\dagger}D^{\dagger} D D i \sigma_3)_{\mu \nu}.
\een
The effect of gauge fixing gives
\ben
Z_W(\{U^s\}) = \int [d (UU^s)]\; W_3(U)\;\Delta_{FP} \; 
\delta[F]\; e^{\beta S(U)},
\een
where we have introduced
\ben
1 = \Delta_{FP}  \int \prod_{j,y} d g_j(y)  \prod_{i,x}
 \delta[F^g_i(U^{\{g_j(y)\}}; x)],
\een
and integrated out the $g$ variables in the standard way.
So $\Delta_{FP} = \det M$ where
\ben
M_{ix;jy} = \left. \frac{\partial F^g_i(x)}{\partial g_j(y)}
\right|_{g=0}
\een

In this case  $Z_W$  is {\em not invariant}.  The shift 
is inconsistent with the gauge condition.   
It is invariant, however, under an infinitessimal shift together 
with an infinitessimal `corrective' gauge transformation
that restores the gauge fixing
\ben
G(x) = \left(1 - \frac{i}{2}\mb{\eta} (x) \cdot  \mb{\sigma} \right)
; \;\;\;\;
U^s(x_0) = \left(1 - \frac{i}{2}\epsilon_{3}(x_0) \sigma_3 \right),
\een
Using the invariance of the measure under combination of
a shift and a corrective gauge transformation we obtain
\ben
\left[ \frac{\partial}{\partial \epsilon_{\mu}(z_0)} + \sum_{k,z} \;\;
\frac{\partial \eta_k(z)}{\partial \epsilon_{\mu}(z_0)}
\frac{\partial }{\partial \eta_k(z)}
\right]Z_W = 0.
\een
In shorthand notation
\cite{dhh98},
the Ehrenfest relation reads
\be
\lla (W_3)_{\mu}\biggr|_s + 
(W_3)_{\mu}\biggr|_g + W_3 \left(
\frac{(\Delta_{FP})_{\mu}}{\Delta_{FP}}\biggr|_s + 
\frac{(\Delta_{FP})_{\mu}}{\Delta_{FP}}\biggr|_g  +
\beta (S)_{\mu}\right)
\rra  = 0.
\ee
Gauge fixing has introduced three new terms:
\begin{itemize}

\item 
$(W_3)_{\mu}\biggr|_g$ comes from the corrective gauge 
transformation acting on the source which is U(1) invariant but
not SU(2) invariant. 

\item 
$\frac{(\Delta_{FP})_{\mu}}{\Delta_{FP}}\biggr|_s$ is the effect of
the shift on the Faddeev-Popov determinant.

\item 
$\frac{(\Delta_{FP})_{\mu}}{\Delta_{FP}}\biggr|_g$ is due to
the corrective gauge transformation of the Faddeev-Popov determinant.

\end{itemize}
The last two derivatives are subtle. The key is to first consider
the constraint up to first order in the shift and the corrective
gauge transformations. Imposing that it is still zero fixes the 
$\{ \eta \}$.
\ben
F_i(x) + \frac{\partial F_i(x)}{\partial \epsilon_\mu(z_0)} 
\epsilon_\mu(z_0)
+ \sum_{k,z} \frac{\partial F_i(x)}{\partial \eta_k(z)}
\eta_k(z) \equiv 0.
\een
Then define the Faddeev-Popov matrix  as a derivative
of the corrected constraint with respect to a general gauge
transformation 
\ben
M_{i x; j y} + \delta M_{i x; j y}  = 
\frac{\partial}{\partial g_j(y)}
\left. \left\{
F^g_i(x) + \frac{\partial F^g_i(x)}{\partial \epsilon_\mu(z_0)} 
\epsilon_\mu(z_0)
+ \sum_{k,z} \frac{\partial F^g_i(x)}{\partial \eta_k(z)} 
\eta_k(z)
\right\} \right|_{g=0}. 
\een
Finally we evaluate the derivative using
\ben
\frac{(\Delta)_{\mu}}{\Delta}  = Tr[ M^{-1} (M)_{\mu}].
\een

\begin{table}[h]
\begin{tabular}{lll} 
\hline \hline
Source:\hspace{4cm} & $W_3$  \hspace{1cm}& $W_3(U \rightarrow D)$ \\ 
Ehrenfest term & &  \\  
\hline \\
$\lla (W_3)_{\mu}\biggr|_s 
\rra $

 &  0.65468(10) \hspace{1cm}&  0.63069(20) \\ \\
 
 $\lla (W_3)_{\mu}\biggr|_g 
\rra$

 &  0.06095(7) \hspace{1cm}&  0.04463(4) \\ \\

$\lla W_3  \frac{(\Delta_{FP})_{\mu}}{\Delta_{FP}}\biggr|_g \rra
$
  
 &  0.00127(21) \hspace{1cm}&  0.00132(50) \\ \\
 
$\lla W_3  \frac{(\Delta_{FP})_{\mu}}{\Delta_{FP}}\biggr|_s \rra
$
  
 &  0.00529(3) \hspace{1cm}&  0.00564(3)  \\ \\
 
$\lla \beta (S)_{\mu}\biggr|_s 
\rra $

 &  -0.72246(68) \hspace{1cm}&  -0.68275(50)   \\ \\ \hline

Zero  & -0.00026(77)& -0.00045(64) \\
\hline \hline
 
\end{tabular}

\caption{Terms in the Ehrenfest relation, Eqn.(1), on a
$4^4$ lattice at $\beta = 2.5$.  The column
labeled $W_3$ corresponds to the source described in the text.
In the second column the source links are replaced by their
diagonal parts
of the links to test a second source. The theorem gives
zero for the sum.}
  
\end{table}

\noindent
A check of the Ehrenfest theorem is given in Table 1. 
Some of the individual terms on the right hand side require a 
$ 2 N \times 2 N $ matrix inversion,  where $N$ is the lattice 
volume.    Hence to test the result 
numerically, we chose as small a lattice as possible. The 
exactness of the theorem
does not involve the size of the lattice which is $4^4$ in Table 1.
The last column employs a different source.  The links making
up the plaquette are replaced by the diagonal parts only as a 
second test of the theorem.

Again by separating the links into diagonal and off-diagonal
parts we get the final form of the Ehrenfest-Maxwell  relation.

\ben
\lla \Delta_{\mu} F_{\mu \nu} \rra = \lla J_{\nu}^{dyn.} \rra
+ J_{\nu}^{static}\biggr|_{s}
+ J_{\nu}^{static}\biggr|_{g}
+ \lla J_{\nu}^{FP}\biggr|_{s}\rra
+ \lla J_{\nu}^{FP}\biggr|_{g}\rra.
 \een
The right hand side consists of a sum of conserved currents.
The first term comes from the excitation of the
charged coset fields,  the static term has an extra non-local
contribution coming from the corrective gauge transformation, 
and the last two contributions are from the ghost fields. 
These terms give a non vanishing charge density cloud around
a static source.  The lefthand side can be used as a lattice
operator to measure the total charge density and does not require
the matrix inversions needed to measure the individual terms 
separately that limited the numerical tests to small lattices.

\begin{table}[h]
\begin{tabular}{cccl} 
\hline \hline
$\beta$ & $div E$(cl.pt.charge) 
 & $div E$ (on source) & total flux  
\\ 
& $= \frac{1}{\beta}$&& \\ \hline \\
10.0 &  0.1 & 0.1042(1) &  0.0910(8) (mid) \\ 
(almost&&& 0.0148(8) (back)\\ 
classical)&&& 0.1092(8) (total)\\

  \\ \hline \\
2.4 &  0.4166 & 0.5385(19) &  0.7455(70) (mid) \\ 
&&& 0.0359(72) (back)\\ 
&&& 0.7815(95) (total)\\ \\ \hline \hline
\end{tabular}

\caption{$div E \equiv \langle \diffn{\nu} F_{\nu 4} \rangle$, 
normalized to $\frac{1}{\beta}$ for a `classical' point charge,
measured on a $3 \times 3$ Wilson loop source on an $8^4$ lattice.  
Integrated electric flux is measured on 
the midplane centered
on the Wilson loop and on a plane on
the far side of the torus, and the sum being the total flux.}

\end{table}

As a simple application we use this definition of flux to
calculate $div E$ on a source and the total flux away from the
source.  In Table 2,  we  see that the total integrated flux on a plane
between the charges plus the flux on a back plane of the torus
is larger than the $div E$ on the source.  The interpretation 
is the  bare charge is dressed with same polarity 
charge by the interactions and the neighborhood has a cloud
of like charge. Hence there is antiscreening.   This charge
density has contributions from all terms in the Ehrenfest relation.
Table 2 also shows that the interactions increase the charge on
the source itself.

\section{Summary}
\label{sec_summ}

We have seen that  the electric charge induced by Abelian
Wilson loop must be reinterpreted. The coset fields renormalise 
the charge of the loop as measured by 
$|\langle \diffn{\nu} F_{\nu \mu} \rangle|$
and charge is also induced in the surrounding vacuum. Full
SU(2) has antiscreening/asymptotic freedom of color
charge, and in the maximal Abelian gauge alone have we seen
analogous behaviour, in that the source charge is increased 
and induces charge of like polarity in the neighboring vacuum.
Whether this renormalization of charge can account for the
reduction of the string tension upon Abelian projection in 
this gauge is not clear.  The improved field strength expression
defined by the Ehrenfest identity does not coincide with the lattice
version
\cite{bdh}
of the 't Hooft field 
strength operator
\cite{thooft74}.

\section*{Acknowledgments}

We wish to thank  G. Di Cecio, A. Di Giacomo, 
G. Gubarev,  M.I. Polikarpov for
discussions.  This work was supported in part by United States 
Department of Energy grant DE-FG05-91 ER 40617.  Y.S. was supported
in part by a Japan Ministry of Education Fellowship.

\end{document}